\documentclass[lettersize,journal]{IEEEtran}
\usepackage{cite}
\usepackage{amsmath,amssymb,amsfonts}
\usepackage{algorithmic}
\usepackage{graphicx}
\usepackage{textcomp}
\usepackage{epstopdf}
\usepackage{multirow}
\usepackage{booktabs}
\usepackage{caption}
\usepackage{footnote}
\usepackage{subfigure}
\usepackage{makecell}
\usepackage{color,bm}
\usepackage{algorithm} 
\newtheorem{prop}{Proposition}
\definecolor{v1}{RGB}{0,0, 0}
\definecolor{v2}{RGB}{0,0,0}
\definecolor{v3}{RGB}{0,0,0}
\definecolor{v4}{RGB}{0,0,0}
\def\BibTeX{{\rm B\kern-.05em{\sc i\kern-.025em b}\kern-.08em
    T\kern-.1667em\lower.7ex\hbox{E}\kern-.125emX}}
\begin{document}
\title{Safe Data-Driven Insulin Dosage Decision Based on Physiology-Constrained Gaussian Process Learning}
\author{\mbox{Deheng Cai, Xiang Lu, Wei Liu, Jing Chen, Linong Ji, Dawei Shi,~\IEEEmembership{Senior Member,~IEEE}}
	\thanks{This work is supported by the National Natural Science Foundation of China under Grants  62403059, 62333001 and 62261160575. (\emph{Corresponding author: Dawei Shi})}
	\thanks{D. Cai is with Machine Learning and I-health International Cooperation Base of Zhejiang Province, School of Automation, Hangzhou Dianzi University, Zhejiang 310018, China. (e-mails: dehengcai@hdu.edu.cn). }
	\thanks{J. Chen, X. Lu, and D. Shi are with MIIT Key Laboratory of Servo Motion System Drive and Control, School of Automation, Beijing Institute of Technology, Beijing, China (e-mails: jingchen@bit.edu.cn, xianglu@bit.edu.cn, daweishi@bit.edu.cn). }
	\thanks{W. Liu and L. Ji are with the Department of Endocrine and Metabolism, Peking University People's Hospital, Beijing, China (e-mails: liuwei850217@163.com, jiln@bjmu.edu.cn). }}
\markboth{IEEE TRANSACTIONS AND JOURNALS TEMPLATE, VOL. XX, NO. XX, November 2026}%
{Shell \MakeLowercase{\textit{et al.}}: A Sample Article Using IEEEtran.cls for IEEE Journals}

\maketitle

\begin{abstract}
Multiple daily injections (MDI) therapy is widely adopted by insulin-treated subjects with diabetes mellitus. In this work, we study how to ensure that data-driven model learning is aligned with the underlying physiological principles, and introduce a safe data-driven learning and decision framework for the MDI therapy. To provide credible glucose predictions for the decision, a physiology-constrained Gaussian process learning method is developed for glycemic dynamics modeling by encoding the physiological priors as constraints in the model training. Considering the natural day-to-day cycle of insulin decisions, a prediction-based optimization formulation with an objective function that has a penalty term associated with the previous dosage is employed for the insulin therapy, together with the asymmetric penalties on hyper- and hypoglycemia for safety. The effectiveness of the proposed method is verified using the 10-adult cohort from the FDA-accepted UVA/Padova T1DM simulator, and an advisory mode analysis is also conducted based on the clinical data from 40 type 1 diabetes and 55 type 2 diabetes subjects to indicate the safe and physiologically plausible decisions in the real-life applications. 
\end{abstract}


\section{Introduction}
\label{sec:introduction}
\IEEEPARstart{D}{iabetes} mellitus (DM) is a persistent metabolic disorder that arises from either a total absence of insulin (type 1 DM, T1DM) or a partial deficiency in insulin production and impaired sensitivity (type 2 DM, T2DM). Individuals with DM frequently encounter long-term complications such as retinopathy and nephropathy, primarily due to inadequate blood glucose (BG) control \cite{diabetes1993effect}. {\color{v1}Currently, insulin therapy with basal-bolus strategy, including multiple daily injections (MDI) and continuous subcutaneous insulin infusions (CSII) via an insulin pump, is commonly adopted for both T1DM and T2DM patients at suitable diabetes courses. In MDI therapy, basal (long-acting) insulin is supplied to maintain normal BG levels during fasting periods and overnight, while bolus (rapid-acting) insulin is administered to address the rise in glucose levels resulting from meal intakes \cite{Thabit2016MDI}.}

{\color{v1}The basal insulin dose is typically determined by clinicians or hospital practitioners based on their clinical judgment;} the meal bolus dosage is computed using a standard calculator based on the carbohydrate content of the meal, the current BG level, and subject-specific settings such as carbohydrate ratio (CR) and correction factor (CF) profiles. Up to now, considerable attention has been paid to improving the performance of the insulin decisions \cite{CAI2023100897}. Given the repetitive nature of insulin decisions, a run-to-run (R2R) control law was developed in \cite{Run2006} to adapt the bolus dosage and timing each day using self-monitoring of blood glucose (SMBG) measurements. In \cite{Cescon2021}, a separate weekly adjustment of bolus and basal dose was investigated utilizing the R2R and iterative learning control, respectively. To exploit more dynamic information contained in continuous glucose monitoring (CGM) data, an R2R-based adaptation framework is proposed in \cite{Herrero2015} for CR and CF based on the minimum postprandial glucose value within a specific time window. Recently, \cite{9195161} considered a model-based R2R method to simultaneously update CR and basal dose daily based on their estimated optimal settings in the previous day. Some efforts have also been made to combine the R2R control with case-based reasoning methods to update CR, CF, and basal dosage \cite{Herrero20151,Herrero2018282,Torrent-Fontbona2019}. Besides, reinforcement learning (RL) methods have become attractive for such subject-specific setting adaptation, such as actor-critic \cite{Sun2019}, double deep Q-learning \cite{10054010} and multi-agent RL \cite{JALOLI2024105762}.

There are also other interesting attempts made to develop data-driven learning and control methods for insulin decisions without relying on patient-specific parameters. For instance, the meal bolus decision is formulated as a risk-sensitive optimization problem in \cite{CAI2021104881} using the predicted glucose trajectory provided by Gaussian process learning. A safe contextual Bayesian optimization algorithm was proposed in \cite{9937070} to iteratively adjust the meal bolus dosage based on the feedback of historical glucose management. In \cite{9308924}, by constructing a penalty cost function using the SMBG measurements, a recursive least square-based extreme seeking control method was investigated for the adaption of basal dosage in T2DM patients. Recently, an optimal impulsive control strategy was considered in \cite{10533199} for insulin decisions, where the periodic glucose trajectory was initially obtained offline under ideal conditions, then tracked in real-time using sparse measurements to account for nonidealities that exist in real life.

Despite considerable research progress, insulin decision methods based on the data-driven model learning can sometimes encounter embarrassing and unsafe decision situations due to the learned input-output relationship that does not conform to physiological laws \cite{prendin2023importance}. {\color{v1}Nowadays, some attempts had been made on integrating soft physiological constraints into kernel-based learning methods \cite{chen2025physiology, feng2025kernel} or utilizing monotonic neural networks \cite{calzavara2026embc} to provide physiologically plausible glucose prediction.} In contrast to prior work, we present an alternative approach that incorporates hard physiological response constraints into Gaussian process learning to directly align the input-output relationships of the model with the physiological response, and further develop a prediction-based optimization framework for insulin dosing decisions. The main contributions are summarized as follows:
\begin{enumerate}
	\item To ensure the physiological relationship between glucose prediction and inputs of insulin dosage and meal intakes, a physiology-informed Gaussian process (GP) learning approach is proposed for glycemic dynamics. Through encoding the physiological effects of insulin and meals on glucose levels as constraints, the hyperparameter is obtained by solving a constrained optimization problem that ensures both the interpretation of the subject-specific glucose data and the enforcement of basic physiological laws.
	\item Based on the credible glucose trajectory prediction, a unified prediction-based optimization framework is developed for both basal and bolus decisions in the insulin therapy, where asymmetric penalties on different glucose situations are designed in the objective function for safety and effectiveness. Besides, a penalty term associated with the previous dosage is incorporated to enable a quick and appropriate decision based on the feedback of previous decision information.
	\item Comprehensive analyses are performed to evaluate the effectiveness of the developed method. Using a 10-adult cohort in the Universities of Virginia (UVA)/Padova T1DM simulator \cite{man2014uva}, the performance of the method is first evaluated using different \emph{in silico} scenarios of potentially incorrect meal carbohydrate counts with nominal and over/under-estimated basal rates. Besides, to verify the potential practical applications of the proposed method, advisory mode analysis is performed based on the clinical data from 40 T1DM and 55 T2DM subjects. 
\end{enumerate}

The remainder of this paper is organized as follows. In Section II, the data-driven insulin decision architecture based on the physiology-constrained learning is introduced. Section III presents the performance evaluation results for both \emph{in silico} simulations and clinical data-based advisory mode analysis. Concluding remarks are outlined in Section IV.

\begin{figure*}[!htb]
	\centering
	\includegraphics[width=\hsize]{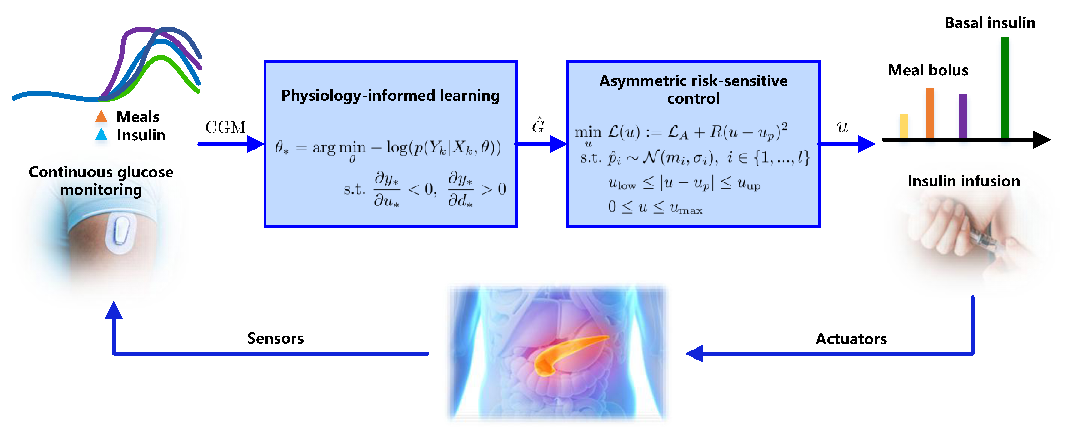}
	\caption{The overall framework of the proposed method. The figures of the subject and physician are from the Internet.}\label{GP-ARS}  
\end{figure*}

\section{Data-driven Insulin Decision for Multiple Daily Injections}\label{section1}
The proposed data-driven decision framework for insulin therapy is illustrated in Fig.~\ref{GP-ARS}. The glucose dynamics are continuously learned using a developed physiology-informed Gaussian process regression method, which can capture the physiological relationship between inputs and glucose prediction. Based on the obtained glucose trajectory predictions, the bolus and basal insulin decisions are then constructed as constrained stochastic optimization problems within the framework of risk-sensitive control.

 {\color{v1}Acknowledging that glucose levels throughout the day are influenced by both the pharmacokinetics (PK) and pharmacodynamics (PD) of basal and bolus insulin, we schedule basal insulin decisions at night and collect night-time glucose data for its model learning. This scheduling minimizes the impact of bolus insulin and daily activities, aligning with clinical practices as described in \cite{liu2024randomized}, which emphasize the importance of night-time glucose levels for basal insulin adjustment.}

\subsection{Gaussian processes}\label{GPML}
A Gaussian process (GP) describes the functions $f(x): \mathbb{R}^n\rightarrow \mathbb{R}$ from a probabilistic perspective, such that values of $f(x)$ at any input $x$ follow a joint Gaussian distribution \cite{Roberts2013, Rasmussen2006}. Generally, the GP is determined by its mean function and covariance function, which have the following forms:
\begin{align}
m(x)&=\mathbb{E}_f[f(x)],\\
k(x, x')&=\text{cov}_f[f(x), f(x')],~x, x'\in\mathbb{R}^n.
\end{align}
Considering observations $y$ of $f(x)$ perturbed by the noise $\omega$ according to $y=f(x)+\omega$, the covariance function of $y$ can be written as:
\begin{align}
k_y(x, x')=k(x,x')+\sigma_\omega^2\delta(x,x'),
\end{align}
where $\omega$ is a Gaussian white noise following $\mathcal{N}(0,\sigma_\omega^2)$ distribution, and $\delta(x, x')$ denotes the Kronecker delta function. Given a dataset $\mathcal{D}:=\{(x_i, y_i)\}_{i=1}^N$, the hyperparameters $\theta$ associated with the mean function and covariance function of the GP are determined through maximizing the log-marginal likelihood: $\mathop{\arg\max}_{\theta} \log(p(Y|X,\theta))$, where $X=[x_1,x_2,...,x_N]^\top$ and $Y=[y_1,y_2,...,y_N]^\top$.

For a new input $x_*$, the trained GP can infer that the prediction output $y_*$ till obeys a Gaussian distribution, whose mean $m_*$ and covariance $\sigma_*^2$ can be calculated as follows:
\begin{align}
m_*&=m(x_*)+K_*^\top(K+\sigma^2_\omega I)^{-1}(Y-m(X)), \label{p1}\\
\sigma_*^2&=k(x_*,x_*)-K_*^\top(K+\sigma^2_\omega I)^{-1}K_*, \label{p2}
\end{align}
where $K_*=[k(x_*,x_1),...,k(x_*,x_N)]^\top$ and $K$ is the covariance matrix having entries $K_{ij}=k(x_i,x_j)$ for $i, j\in \{1, 2, ..., N\}$. Therefore, the prediction and its estimated uncertainty are described by \eqref{p1} and \eqref{p2}, respectively. 

\subsection{Physiology-informed learning of glucose dylnamics}\label{ESO}
In general, ML-based algorithms for BG prediction are black-box modeling methods, which lack the interpretability of the outcome to get physicians' favor for clinical use \cite{prendin2023importance}. Besides, the non-physiologically sound prediction methods integrated in decision support systems may lead to the inappropriate direction for dosage adjustment, increasing the risks of hypoglycemia or hyperglycemia for patients. 
In this regard, we here present an attempt at physiology-informed learning of glucose dynamics based on the GP regression method. 

Specifically, the main goal of the GP modeling is to describe the glucose dynamics after bolus and basal insulin, thus providing long-term prediction information for the insulin decision. In this context, the sampled CGM data for the insulin event $n$ are divided into the glucose measurement sequence before and after insulin injection, which are defined as $G_{pre}$ and $G_{post}$, respectively. In detail, $G_{pre} = [p_1,\cdots,p_m]$ includes $m$ glucose data points $p$, and $G_{post} = [p_{m+1},\cdots,p_{m+l}]$ has $l$ glucose data points $p$, which are selected to have an equal time distance $\Delta t$. {\color{v1}The parameters $m, l$ and $\Delta t$ are determined based on the different prediction horizon requirements for basal and bolus insulin decisions. }
With the sequences $G_{pre}$ and $G_{post}$, one step-ahead strategy is further adopted for the GP modeling by defining the following inputs $x_i^n$ and outputs $y_i^n$:
\begin{align}
x_{i}^n=&[p_{i},\cdots,p_{m+i-1},u_n,d_n]^\top,\nonumber\\
y_{i}^n=&p_{m+i} - p_{m+i-1},~i=\{1,2,\cdots,l\},
\label{model}
\end{align}
where $u_n$ denotes the insulin dosage of the event $n$, and $d_n$ is the carbohydrate (CHO) intake only for the bolus insulin event. Note that the output is designed in the form of difference to reduce the prediction uncertainty \cite{Deisenroth2015}. 
Moreover, $u_n$ and $d_n$ stay the same for the samples from the same event $n$. 
To highlight the effects of $u_n$ and $d_n$ on glucose regulation, the normalization procedure is performed for the glucose measurements in the inputs.

Denoting the available dataset for current insulin decision as $\mathcal{D}_k:=\{(x_i^n, y_i^n)_{i=1}^l\}_{n=1}^k$, the classic log-marginal likelihood for the hyperparameter determination has the following form:
\begin{align}
\log(p(Y_k|X_k,\theta)) =& -\frac{1}{2}\bar{Y}_k^\top(K(X_k, X_k)+\sigma_\omega^2I)^{-1}\bar{Y}_k\\&-\frac{1}{2}\log|K(X_k, X_k)+\sigma_\omega^2I|-\frac{N}{2}\log2\pi,\nonumber
\end{align}
where $\bar{Y}_k = Y_k - m(X_k)$, and $X_k$ and $Y_k$ is stacked by $x_i^n$ and $y_i^n$ from the dataset $\mathcal{D}_k$, respectively. Due to the complexity of the glucose metabolism, the determined hyperparameter may correspond to a particular interpretation of the data, {\color{v1}but dose not follows} the basic physiological priors of glucose dynamics in the prediction, e.g., insulin delivery leads to a descent of glucose while carbohydrate intake causes an elevation in glucose levels. Therefore, a physiology-informed GP learning method is developed here by encoding the above physiological priors as constraints to the hyperparameter optimization, which is in the following form:
\begin{align}
\theta_* = \arg \min_{\theta}& -\log(p(Y_k|X_k,\theta)), \label{hyperparameter}\\
\text{s.t.}&~\frac{\partial y_*}{\partial u_*} < 0, ~\frac{\partial y_*}{\partial d_*} > 0, \label{constrains}
\end{align}
where $y_*$ is the predicted mean corresponding to the insulin dosage $u_*$ and carb intake $d_*$. The specific forms of the constraints depend on the designed mean and covariance functions of the GP. 

In this work, the squared exponential (SE) covariance kernel is used as the covariance function, and the mean function is designed in a linear form:  
\begin{align}
m(x_i)=&a^\top x_i+b,\label{mean}\\
k(x_i, x_i')=&\sigma_f^2\exp\left(-\frac{1}{2}(x_i-x_i')^\top\Omega^{-1}(x_i-x_i')\right),\label{SE}
\end{align} 
where $a=[a_{1},\cdots,a_{m},a_u,a_d]^\top$, $\Omega=\text{diag}\{[l_1^2, l_2^2,...,l_{m+2}^2]\}$, and $\sigma_f^2$ denotes the signal variance. With the above mean and covariance functions, the constraints in \eqref{constrains} can be analytically constructed, which are shown in the following proposition. 
\begin{prop}
With the mean and covariance functions of GP designed in \eqref{mean}-\eqref{SE}, the constraints $\frac{\partial y_*}{\partial u_*} < 0$ and $\frac{\partial y_*}{\partial d_*} > 0$ are equivalent to $a_u <-\sqrt{\sum_{i=1}^N \alpha_i^2\sum_{i=1}^N (\sigma_f^2\frac{\exp(-\frac{1}{2})}{|l_{m+1}|})^2}$ and $a_d > \sqrt{\sum_{i=1}^N \alpha_i^2\sum_{i=1}^N (\sigma_f^2\frac{\exp(-\frac{1}{2})}{|l_{m+2}|})^2}$, where $\alpha_i$ is the $i$th element of $\alpha := (Y_k-m(X_k))^{\top}(K+\sigma^2_\omega I)^{-1}$.
\end{prop}
\begin{IEEEproof}
See Appendix A.
\end{IEEEproof}

Finally, the interior point method is applied to solve the problem in \eqref{hyperparameter} and \eqref{constrains}. 
After model training, the predicted long-term glucose trajectory $\hat{G}=[\hat{p}_1, \hat{p}_2, ... ,\hat{p}_l]^\top$ is obtained by successively feeding back the predicted glucose mean from the previous step into the input for current prediction. 

\subsection{Asymmetric risk-sensitive control for insulin decision}\label{NF}
Based on the glucose prediction information $\hat{G}$, the insulin decision is formulated as a prediction-based optimization problem. Specifically, considering the asymmetric risks of hypoglycemia and hyperglycemia, as well as the uncertainty in glucose prediction, an asymmetric cost function is developed within the risk-sensitive (RS) framework \cite{Whittle1986}: 
\begin{align}
\mathcal{L}_{A}=-\frac{2}{\gamma}\log\mathbb{E}\left[\exp\left(-\frac{\gamma}{2}(\hat{G}-G_r)^\top Q(\hat{G}-G_r)\right)\right],\label{ARS}
\end{align}
where $G_r$ is a personalized glucose target for the bolus or basal insulin decision; $Q$ is an asymmetric penalty matrix in the diagonal form; $\gamma<0$ is a risk sensitivity parameter that determines the system's attitude towards uncertainty, and $\gamma<0$ tends to make the optimizer choose the control actions that reduce the prediction uncertainty \cite{Yang2015,Pan2018}. 

Similar to the work in \cite{CAI2021104881}, $Q$ is designed in the following form to enforce the asymmetric penalty on the glucose excursions above and below the target:
\begin{align}
Q_i&=\left\{\begin{array}{ll}
Q_b^i\left(\frac{c_1}{1+\exp\{\alpha (\beta-|\hat{p}_i-g_{i}|)\}}+c_2\right), &\textrm{if}~e_i > 0,\\
Q_b^i,&\textrm{otherwise},
\end{array}\right.
\end{align}
where $e_i=\mathbf{1}(\hat{p}_i-g_{i}<0)$, $g_{i}$ is the $i$th element of $G_r$ and $\mathbf{1}(\cdot)$ denotes the indicator function; $Q_i$ is the $i$th diagonal element of $Q$, and $Q_b^i > 0$ is a basic penalty value; $\Gamma := [\alpha,\beta,c_1,c_2]$ is a quadruple determines the penalty intensity, and is designed to produce a larger penalty on the glucose excursions below the target, so as to take extra care of hypoglycemia.

The insulin decision for the therapy is characterized by the natural day-to-day cycle, where the historical insulin dosage and its glucose management performance would provide informative support for the current insulin decision. In this regard, given the designed asymmetric RS cost $\mathcal{L}_{A}$ in \eqref{ARS}, the stochastic optimization problem in \cite{CAI2021104881} is improved for the insulin decision through incorporating a penalty term associated with the previous insulin dosage:
\begin{align}
\min_{u}&~\mathcal{L}(u):= \mathcal{L}_{A}+R(u-u_p)^2,\label{NOP}\\
\text{s.t.}& ~\hat{p}_i\sim\mathcal{N}(m_i,\sigma_i),~i\in\{1,...,l\},\\
&~u_{\text{low}} \leq |u - u_p| \leq u_{\text{up}},\label{bound1}\\
&~0 \leq u \leq u_{\text{max}},\label{bound}
\end{align}where $u$ is the insulin dosage to be optimized, and $u_p$ denotes the previous insulin dosage for the same meal type (breakfast, lunch, and dinner) or basal titration. $R$ is a weight to balance the belief in the historical glucose management result and the current prediction-based glucose control. $m_i$ and $\sigma_i$ are parameters provided by the GP prediction. Since the gradient of the cost function cannot be obtained analytically, Bayesian optimization (BO) \cite{Shahriari2015T} is applied to solve the above constrained stochastic optimization problem, in which the value of the cost function in \eqref{NOP} is estimated by Monte-Carlo simulations. 
The detailed solution steps can be found in \cite{CAI2021104881}. To ensure safety, the final meal bolus is  constrained by an insulin on board (IOB) term $u_{\text{IOB}}$ \cite{Gondhalekar2018} to prevent over-delivery.
\begin{table}[!htb]
	\caption{Parameters for the proposed method}
	\centering
	\resizebox{\columnwidth}{!}{
		\begin{tabular}{lll}  
			\hline
			\hline
			Variables&description&Value\\\hline 
			\multirow{2}{*}{$m$}&Number of glucose points&\multirow{2}{*}{8~(bolus), 4~(basal)} \\&before insulin injection&\\
			\multirow{2}{*}{$l$}&Number of glucose points&\multirow{2}{*}{8~(bolus), 9~(basal)} \\&after insulin injection&\\
			\multirow{2}{*}{$\Delta t$}&\multirow{2}{*}{Time interval of glucose points}&15 [min] (bolus)\\ &&60 [min] (basal)\\
			$\gamma$&Risk sensitivity parameter&-2\\
			$R$&Input weight&20~(bolus), 10~(basal)\\
			$u_{\text{max}}$&Maximum input constraint&20~(bolus), 30~(basal)\\
			$u_{\text{low}}$&Lower bound of adjustment&3\\
			$u_{\text{up}}$&Up bound of  adjustment&3\\
			\multirow{2}{*}{$Q_b$}&\multirow{2}{*}{Basic penalty matrix}&$\text{diag}\{[0.0025,0.0025,$\\&&$\cdots,0.005,0.005]\}$\\
			$\Gamma$&Quadruples for insulin decision&$[1,10,4,1]$\\
			\multirow{4}{*}{$G_r$}&\multirow{2}{*}{Glucose Target for bolus decision}&$[100,120,140,160,$\\&&$160,150,140,140]^\top$\\&\multirow{2}{*}{Glucose Target for basal decision}&$[80, 80, 100, 100, 110,$\\&&$110, 130, 130, 130]^\top$\\
			\hline
			\hline
	\end{tabular}}\label{T1}
\end{table}
\section{Performance Analysis}\label{spa}
The performance of the proposed method is evaluated on the UVA/Padova T1DM simulator using the 10-adult cohort, and also validated based on advisory-mode analysis \cite{Gillis2007} using the clinical data from T1DM and T2DM subjects who received MDI therapy.  
\subsection{Experiment setup}\label{spa}
In the \emph{in silico} simulations, a four-day (96-hour) protocol begins at 05:00 on the first day with different scenarios employed to evaluate the effectiveness and robustness of the proposed method. Specifically, breakfast, lunch, and dinner are provided every day at the time randomly drawn from [07:00, 09:00], [11:00, 13:00], and [18:00, 20:00], respectively, and the meal sizes obey the normal distribution with means and standard deviations equal to [50, 75, 75] g and [3, 4, 4] g CHO. {\color{v1}Besides, meal bolus is determined with incorrect CHO counts to simulate the inevitable deviations in real life. According to the reports of the existing literature \cite{brazeau2013carbohydrate, meade2016accuracy}, the estimation errors of most patients in CHO counting is typically between 20\% and 40\%. In this way, the counting errors are assumed to be uniformly distributed in [-30\%, +30\%]. 
	
In the first two days of the protocol, the meal bolus is decided by the standard bolus calculator commonly adopted in clinical practice \cite{2008Benefits, Klonof2012The}, which is in the form of
\begin{align}
u_{\text{bolus}} = \frac{\text{CHO}}{\text{CR}}+\frac{G_{c}-G_{sp}}{\text{CF}}-u_{\text{IOB}},
\end{align}}where $G_{c}$ denotes the current glucose level (mg/dL); $G_{sp}$ is the glucose target (mg/dL), 
and selected as 140 for the all virtual patients. In the following two days, the bolus is determined by the proposed method and also compared to the bolus calculator. {\color{v1}Note that since the exact values of CR and CF are known for the simulator, the meal bolus calculator in (20) would virtually provide the ``optimal'' performance and is consequently utilized as the benchmark for comparison.} The model learning is performed using the samples collected in the first two days and continuously updated using the newly generated samples after every decision. 
\begin{table*}[t]
	\centering
		\caption{Comparison results for the scenarios of nominal basal rate}
		\label{T3}
		\begin{tabular}{lcclccl}
			\hline
			\hline\noalign{\vspace{2pt}}
			Scenario (\#Simulations=10)&\multicolumn{3}{c}{Regular meals}&\multicolumn{3}{c}{Irregular meals}\\\cmidrule(lr){2-4}\cmidrule(lr){5-7} Metric &Control &Proposed &$p$ value &Control  &Proposed &$p$ value \\ \hline
			\% time&\\
			~~~~~~$<$54 mg/dL& 0.0 (0.0) &0.0 (0.0) &1.000& 0.0 (0.0) &0.0 (0.0) &1.000\\
			~~~~~~$<$70 mg/dL& 0.0 (0.0) &0.0 (0.0) &0.500& 0.0 (0.2) &0.0 (0.0) &0.625\\
			~~~~~~70-180 mg/dL& 85.9 (8.7) &\textbf{90.1} \textbf{(4.7)} &\textbf{0.002}& 86.1 (5.7) &\textbf{89.8} \textbf{(2.6)} &\textbf{0.004}\\
			~~~~~~$>$180 mg/dL& 14.1 (8.7) &\textbf{9.4} \textbf{(4.7)} &\textbf{0.004}& 13.8 (5.7) &\textbf{10.0} \textbf{(2.6)} &\textbf{0.004}\\
			~~~~~~$>$250 mg/dL& 0.0 (0.0) &0.0 (0.0) &1.000& 0.0 (0.0) &0.0 (0.0) &1.000\\
			Mean glucose(mg/dL)& 136.6 (13.9) &\textbf{130.2} \textbf{(2.8)} &\textbf{0.006}& 138.7 (13.5) &\textbf{129.7} \textbf{(11.9)} &\textbf{0.006}\\
			SD glucose(mg/dL)& 32.1 (10.1) &\textbf{29.3} \textbf{(9.2)} &\textbf{0.002}& 35.8 (7.6) &\textbf{31.2} \textbf{(7.6)} &\textbf{0.002}\\
			Mean glucose at 07:00 (mg/dL)& 121.9 (3.5) &\textbf{121.1} \textbf{(3.9)} &\textbf{0.004}& 122.5 (4.4) &\textbf{121.7} \textbf{(4.5)} &\textbf{0.002}\\
			\hline
			\multicolumn{7}{l}{Data in this table are shown as median (inter quartile range). $p$ values are calculated based on the number of}\\
			\multicolumn{7}{l}{virtual patients, and Statistically significant ($p < 0.05$) changes are highlighted in bold.}
	\end{tabular}
	\end{table*}
	
\begin{figure*}[!htb]
\centering
\subfigure[Regular meal intakes]
{\includegraphics[width=0.49\textwidth]{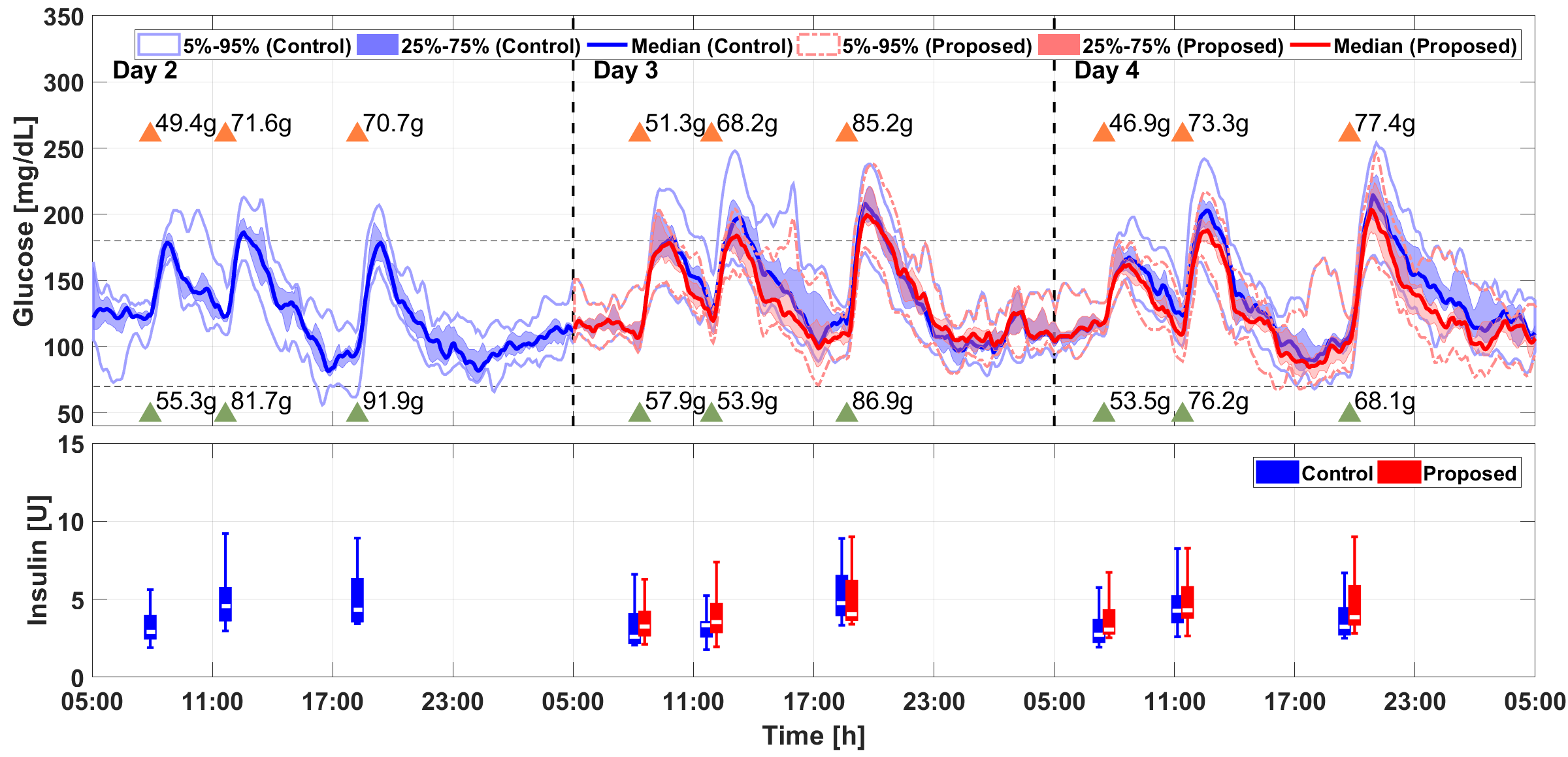}\label{nominal_basal}}
\subfigure[Irregular meal intakes]
{\includegraphics[width=0.49\textwidth]{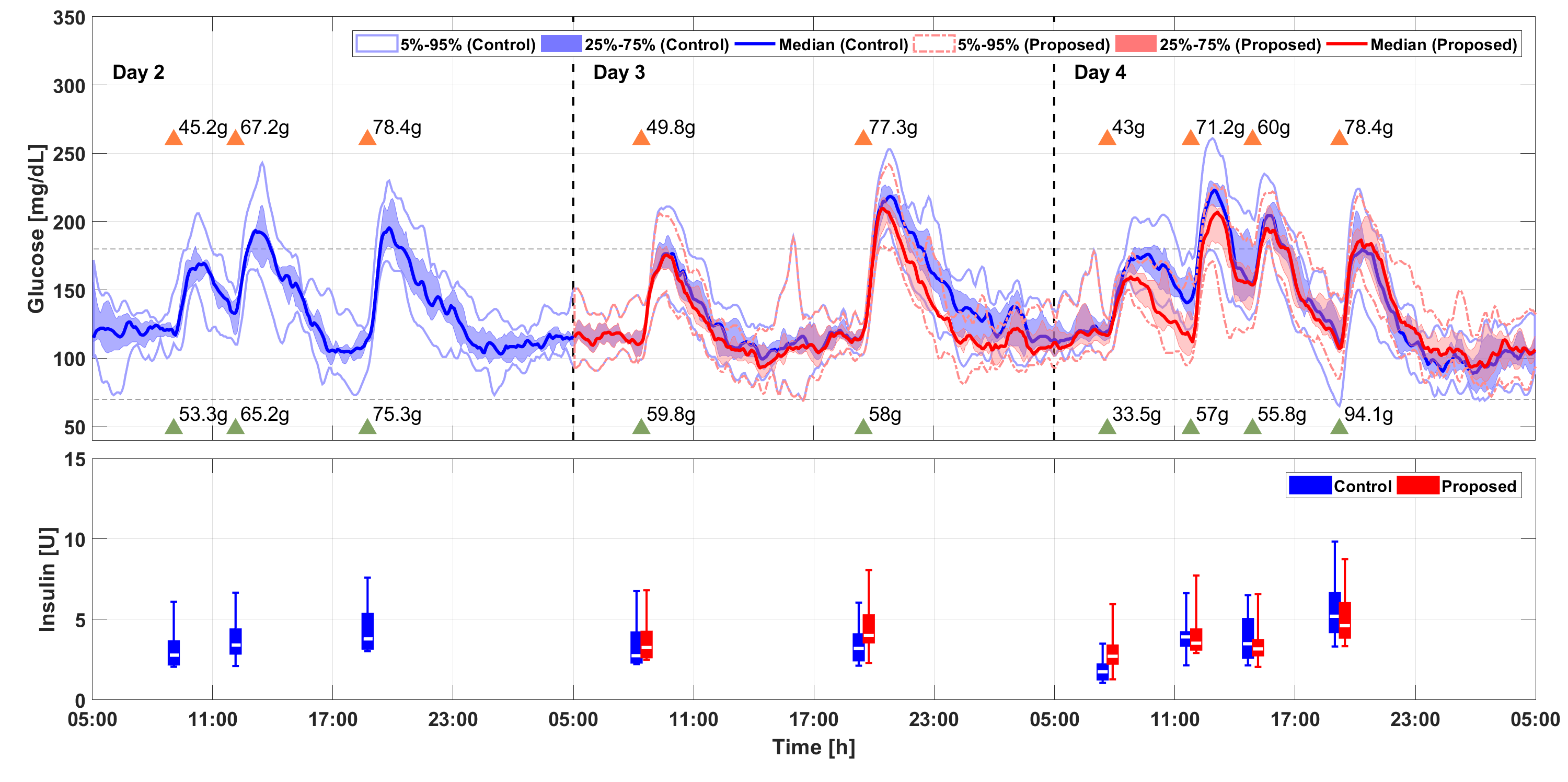}\label{MultMeal}}
\caption{Performance comparison with nominal basal rates in terms of glucose regulation and insulin decision. Orange and green triangles denote actual and estimated meal sizes, respectively.}
\label{nbasal}
\end{figure*}
{\color{v4}Since there is no action model of long-acting basal insulin in the simulator, we focus on the performance evaluation of the proposed method in meal bolus decision to demonstrate the feasibility of the proposed data-driven prediction-based optimization framework for the insulin decision.} The parameters selected by the proposed method are provided in Table \ref{T1}. The simulation is performed on the 10-adult cohort using the above protocol for scenarios of nominal basal rate and its mismatches (110\% and 80\% of the nominal basal rate) with additive CGM noises, thus a total of 10 simulations are performed for each scenario. 
The statistical results with nominal basal rate are presented as ``Regular meals'' in Table~\ref{T3}, and the results with basal rate mismatch are shown in Table~\ref{T4}. The corresponding graphical illustrations are provided in Figs.~\ref{nbasal}-\ref{dbasal}. 
		
To show the robustness of the proposed method in irregular meal situations, the performance of the method is evaluated by considering the scenario of nominal basal rate with irregular meal intakes. Specifically, a four-day protocol beginning at 05:00 on the first day is employed, and the dietary conditions are the same as the previous protocol except for a skipped lunch meal in the third day and an additional meal intake happened around 15:00 on forth day. The results for the last two days are also compared with the standard bolus calculator. The statistical results are presented as ``Irregular meals'' in Table~\ref{T3}, and the corresponding graphical illustrations are provided in Fig.~\ref{MultMeal}.
Following the principles of advisory-mode analysis \cite{Gillis2007}, the glucose data obtained in the clinical trial are fed to the proposed method, thus performing insulin decisions, which are then compared with the historical dosages suggested by clinicians. The corresponding results are presented in Table~\ref{T7}, and the graphical illustrations to the glucose predictions are provided in Fig.~\ref{uob}. To further illustrate the advantages of the proposed method equipped with the physiology-informed predictor, using the typical examples from the clinical data, a comparison of insulin decision is conducted between the proposed method with and without utilizing glucose metabolism priors in modeling, which can be regarded as a basic on-line data-driven prediction-based decision method. The corresponding graphical illustrations are provided in Fig.~\ref{examples}.
\subsection{Various basal rate conditions}\label{A}
{\color{v1}In this subsection, the simulation results for the cases with nominal basal rate and its mismatches are presented. For the scenario of nominal basal rate, from Table~\ref{T3}, the proposed method achieves slightly improved glucose regulation performance in comparison with the standard insulin bolus calculator, which is reflected in percent time in the euglycemic range of 70-180 mg/dL (90.1\% vs. 85.9\%, $p=0.002$), mean glucose (130.2 mg/dL vs. 136.6 mg/dL, $p=0.006$). No hypoglycemia events are observed (percent time $<$ 70 mg/dL, 0.0\% vs. 0.0\%,  $p=0.500$). The statistical results are consistent with the quartile curves in Fig.~\ref{nominal_basal}.} 
\begin{table*}[t]
	\centering
		\caption{Comparison results for the scenarios of under/over-estimated basal rate}
		\label{T4}
		\begin{tabular}{lcclccl}
			\hline
			\hline\noalign{\vspace{2pt}}
			Scenario (\#Simulations=10)&\multicolumn{3}{c}{ 80\% of the nominal basal rate}&\multicolumn{3}{c}{110\% of the nominal basal rate}\\\cmidrule(lr){2-4}\cmidrule(lr){5-7} Metric &Control &Proposed &$p$ value &Control  &Proposed &$p$ value \\ \hline
			\% time&\\
			~~~~~~$<$54 mg/dL& 0.0 (0.0) &0.0 (0.0) &1.000& 0.0 (0.0) &0.0 (0.0) &1.000\\
			~~~~~~$<$70 mg/dL& 0.0 (0.0) &0.0 (0.0) &1.000& 0.5 (3.5) &0.3 (0.9) &0.156\\
			~~~~~~70-180 mg/dL& 71.7 (10.8) &\textbf{84.7} \textbf{(4.9)} &\textbf{0.002}& 89.8 (7.8) &89.4 (4.5) &0.922\\
			~~~~~~$>$180 mg/dL& 28.3 (10.8) &\textbf{15.3} \textbf{(4.9)} &\textbf{0.002}& 9.2 (8.0) &10.6 (4.3) &0.193\\
			~~~~~~$>$250 mg/dL& 0.0 (1.7) &0.0 (0.0) &0.125& 0.0 (0.0) &0.0 (0.0) &1.000\\
			Mean glucose(mg/dL)& 161.6 (8.6) &\textbf{146.5} \textbf{(12.7)} &\textbf{0.002}& 121.8 (12.7) &\textbf{130.1} \textbf{(11.8)} &\textbf{0.037}\\
			SD glucose(mg/dL)& 30.1 (8.5) &28.6 (7.7) &0.131& 34.6 (7.5) &36.5 (6.7) &0.193\\
			Mean glucose at 07:00 (mg/dL)& 162.0 (5.0) &\textbf{158.0} \textbf{(6.5)} &\textbf{0.002}& 100.3 (4.3) &100.1 (5.1) &0.262\\
			\hline
			\multicolumn{7}{l}{Data in this table are shown as median (inter quartile range). $p$ values are calculated based on the number of}\\
			\multicolumn{7}{l}{virtual patients, and Statistically
				significant ($p < 0.05$) changes are highlighted in bold.}
	\end{tabular}
\end{table*}
\begin{figure*}[!htb]
	\centering
	\subfigure[Under-estimated basal rates]
	{\includegraphics[width=0.49\textwidth]{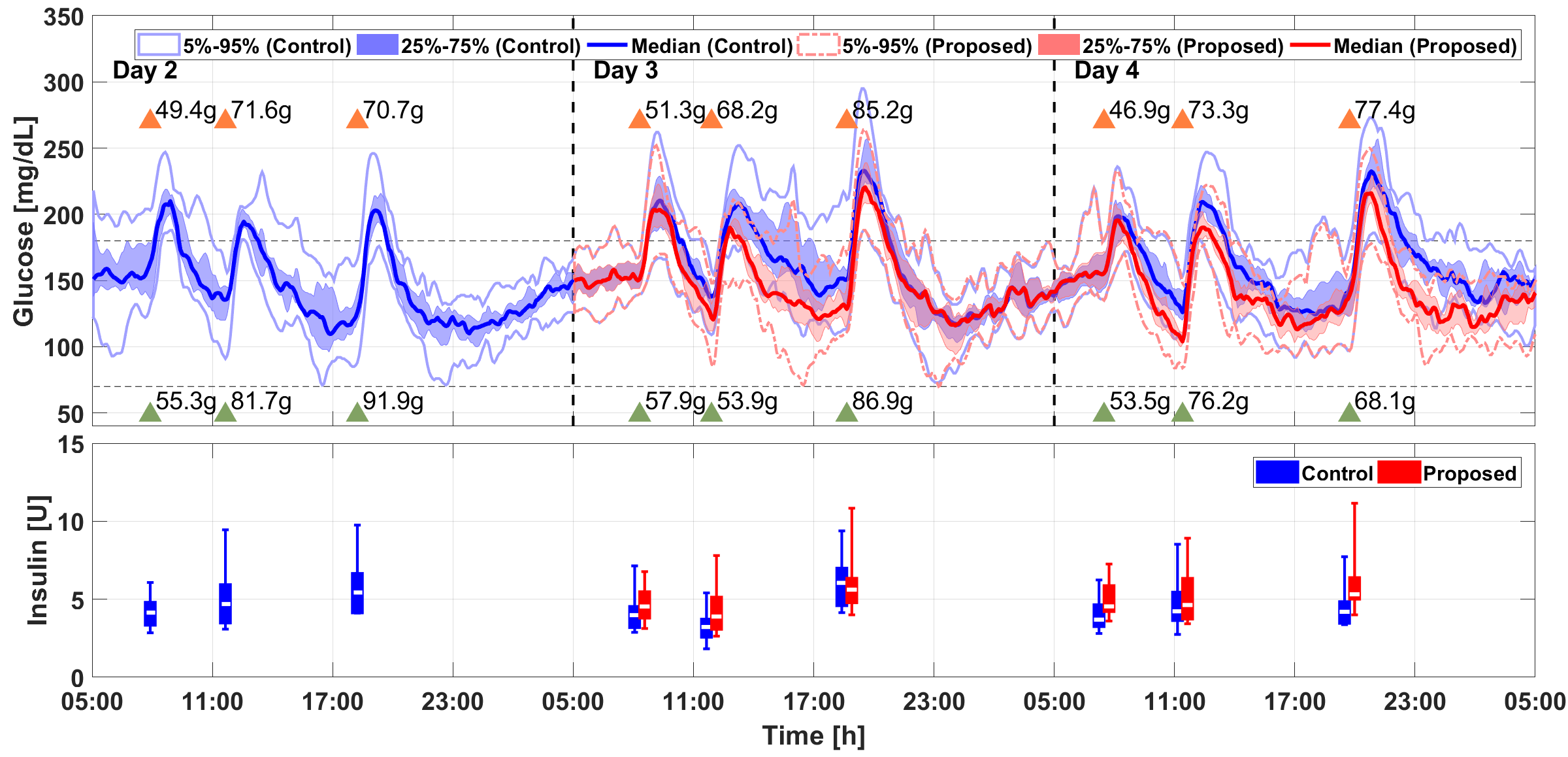}\label{08basal}}
	\subfigure[Over-estimated basal rates]
	{\includegraphics[width=0.49\textwidth]{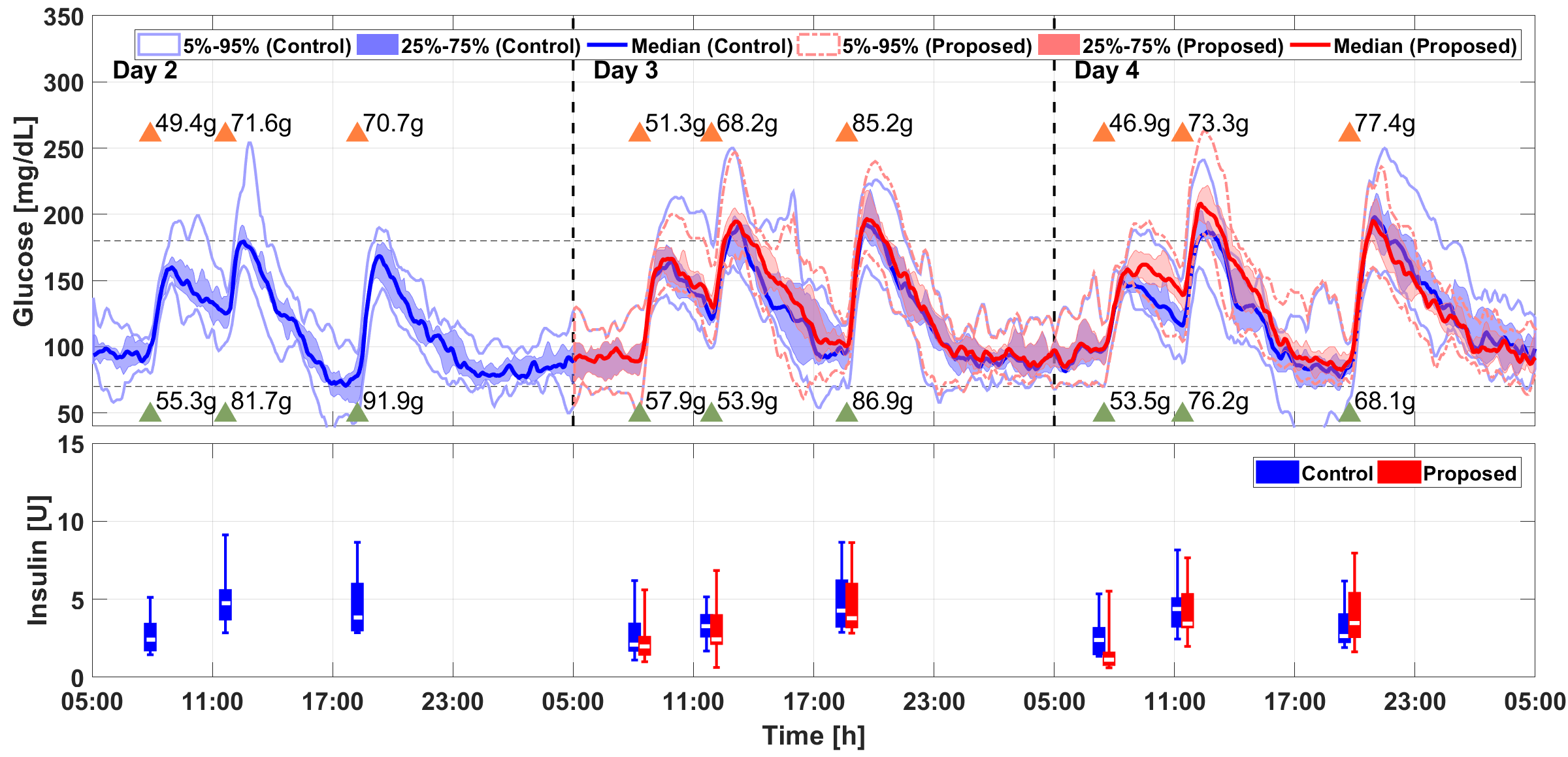}\label{12basal}}
	\caption{Performance comparison with different basal rates in terms of glucose regulation and insulin decision. Orange and green triangles denote actual and estimated meal sizes, respectively.}
	\label{dbasal}
\end{figure*}

For the scenario of under-estimated basal rate, it is observed from Fig.~\ref{08basal} that compared with the standard bolus calculator (using near-optimal CR and CF values provided by the simulator), the proposed method tends to adopt more responsive but safe actions for the elevated glucose levels at day 3 and 4, thus achieves improved glucose regulation performance for both bolus and basal insulin decisions (see Table~\ref{T4}). This is reflected in percent time in the euglycemic range of 70-180 mg/dL (71.7\% vs. 84.7\%, $p=0.002$), mean glucose (146.5 mg/dL vs. 161.6 mg/dL, $p=0.002$). No hypoglycemia events are observed (percent time $<$ 70 mg/dL, 0.0\% vs. 0.0\%,  $p=1.000$).

Glucose management is threatened by a greater risk of hypoglycemia in the scenario of over-estimated basal rate. Compared with the bolus calculator, the proposed method exhibits more conservative decision actions, which reduce risks of hypoglycemia and more elevated postprandial glucose levels at days 3 and 4 in Fig.~\ref{12basal}. These observations are supported by the comparison results in Table~\ref{T4}, where the proposed method achieves tighter and safer glucose control in terms of slightly reduced percent time below 70 mg/dL (0.3\% vs. 0.5\%, $p = 0.156$) and similar percent time in [70, 180] mg/dL (89.4\% vs. 89.8\%, $p$ = 0.922) while reasonably increasing mean glucose (130.1 mg/dL vs. 121.8 mg/dL, $p = 0.037$). These results indicate the proposed method is able to perform reasonable insulin decisions, and is also robust to both meal size counting disturbances and basal rate mismatches. 

\subsection{Irregular meal intakes}
{\color{v1}In this subsection, the simulation results for the scenario of nominal basal rate with irregular meal intakes are presented. We can observe from Fig.~\ref{MultMeal} that despite the occurrence of skipped lunch meal in day 3 and multiple meals in day 4, the proposed method achieves satisfactory glucose management performance, which is supported by the statistical results in Table~\ref{T3}. Specifically, compared with the standard bolus calculator, a slightly improved results are obtained by the proposed method in terms of improved percent time in [70, 180] mg/dL (86.1\% vs. 89.8\%, $p$ = 0.004), reduced mean glucose (138.7 mg/dL vs. 129.7 mg/dL, $p = 0.006$) and reduced hyperglycemia events, defied as percent time above 180 mg/dL (13.8\% vs. 10.0\%, $p$ = 0.004). }
\subsection{Advisory-mode evaluations on clinical data}\label{C}
To evaluate the potential practical applications of the proposed method, the historical clinical data from both T1DM and T2DM subjects who undertook MDI therapy are included in the advisory mode analysis. Intermittently scanned CGM (Freestyle Libre H, Abbott, US) was worn to monitor subcutaneous interstitial glucose. Subjects were provided with three meals per day during hospitalization, which were managed to have similar meal timing and sizes. The meal bolus and basal insulin were determined by clinicians according to current clinical strategies based on intermittently scanned CGM (isCGM) readings. Besides, basal insulin injection is scheduled around 10 pm every night. The studies for T2DM and T1DM subjects were approved by the institutional review board at Peking University People's Hospital (2020PHB338-01 and 2022PHB407-001), respectively, and written informed consents of all participants were obtained. {\color{v1}The study of 2020PHB338-01 was a randomized clinical trial aiming to evaluate the safety and efficiency of intelligent meal bolus insulin dosage decision algorithm in T2DM patients  with MDI therapy \cite{liu2024randomized}, while the study of 2022PHB407-001 was an observational study on T1DM patients with MDI therapy.}
\begin{table}[t]
	\caption{Demographic characteristics of T1DM and T2DM subjects}
	\centering
	\resizebox{\columnwidth}{!}{
		\begin{tabular}{lcc}  
			\hline
			\hline\noalign{\vspace{2pt}}
			&T1DM (n = 40)&T2DM (n = 55)\\\hline \noalign{\vspace{2pt}}
			Gender (male/female)&19/21&36/19\\
			Age (years)  &49.2 $\pm$ 19.9&60.5 $\pm$ 13.6\\
			Age at diagnosis (years)&46.9 $\pm$ 18.6&46.3 $\pm$ 13.4\\
			Diabetes duration (years) &12.4 $\pm$ 9.7&14.0 $\pm$ 9.1\\
			Weight (kg) &62.7 $\pm$ 10.7&73.9 $\pm$ 11.1\\
			Height (cm)&163.0 $\pm$ 9.6&165.2 $\pm$ 9.1\\
			BMI (kg/m2)&23.6 $\pm$ 2.9&27.1 $\pm$ 3.5\\
			SBP  (mmHg)  &129.0 $\pm$ 16.4&138.0 $\pm$ 16.9\\
			DBP  (mmHg)  &72.4 $\pm$ 10.2&78.7 $\pm$ 11.9\\
			HbA1c (\%)   &9.3 $\pm$ 1.9 &10.2 $\pm$ 1.8\\\noalign{\vspace{2pt}}
			\hline
			\multicolumn{3}{l}{Data are mean $\pm$ standard deviation. Abbreviations: BMI, body}\\
			\multicolumn{3}{l}{mass index; SBP, systolic blood pressure; DBP, diastolic blood}\\
			\multicolumn{3}{l}{pressure; HbA1c, haemoglobin A1c.}
	\end{tabular}}\label{T5}
\end{table}	
\begin{table}[t]
	\centering
		\caption{Evaluation results on clinical data for the proposed method}
		\begin{tabular}{cccc}
			\hline
			\hline\noalign{\vspace{2pt}}
			Dataset& Metric&Bolus decision&Basal decision\\
			\hline\noalign{\vspace{2pt}}
			\multirow{4}{*}{T1DM}&MAE (U)& 1.24 (1.15, 1.34) &1.46 (1.29, 1.62) \\
			&CA (\%)&  65.1 (328/504) & 83.6 (138/165) \\
			&IA (\%)& 28.0 (141/504) & 18.2 (30/165) \\
			\hline\noalign{\vspace{2pt}}
			\multirow{4}{*}{T2DM}&MAE (U)& 1.15 (1.03, 1.27) & 1.42 (1.15, 1.69) \\
			&CA (\%)& 73.2 (262/358) & 85.6 (107/125)\\
			&IA (\%)& 34.1 (122/358) & 40.8 (51/125)\\
			\hline\noalign{\vspace{2pt}}
			\multicolumn{4}{l}{Data in this table are shown as mean (95\% CIs) or \% (n/N).} \\
			\multicolumn{4}{l}{N denotes the decision counts for the T1DM or T2DM dataset.} 
	\end{tabular}\label{T7}
	\end{table}
	\begin{figure*}[!htb]
		\centering
		\centering
		\subfigure[A patient from the T1DM dataset]
		{\includegraphics[width=0.49\textwidth]{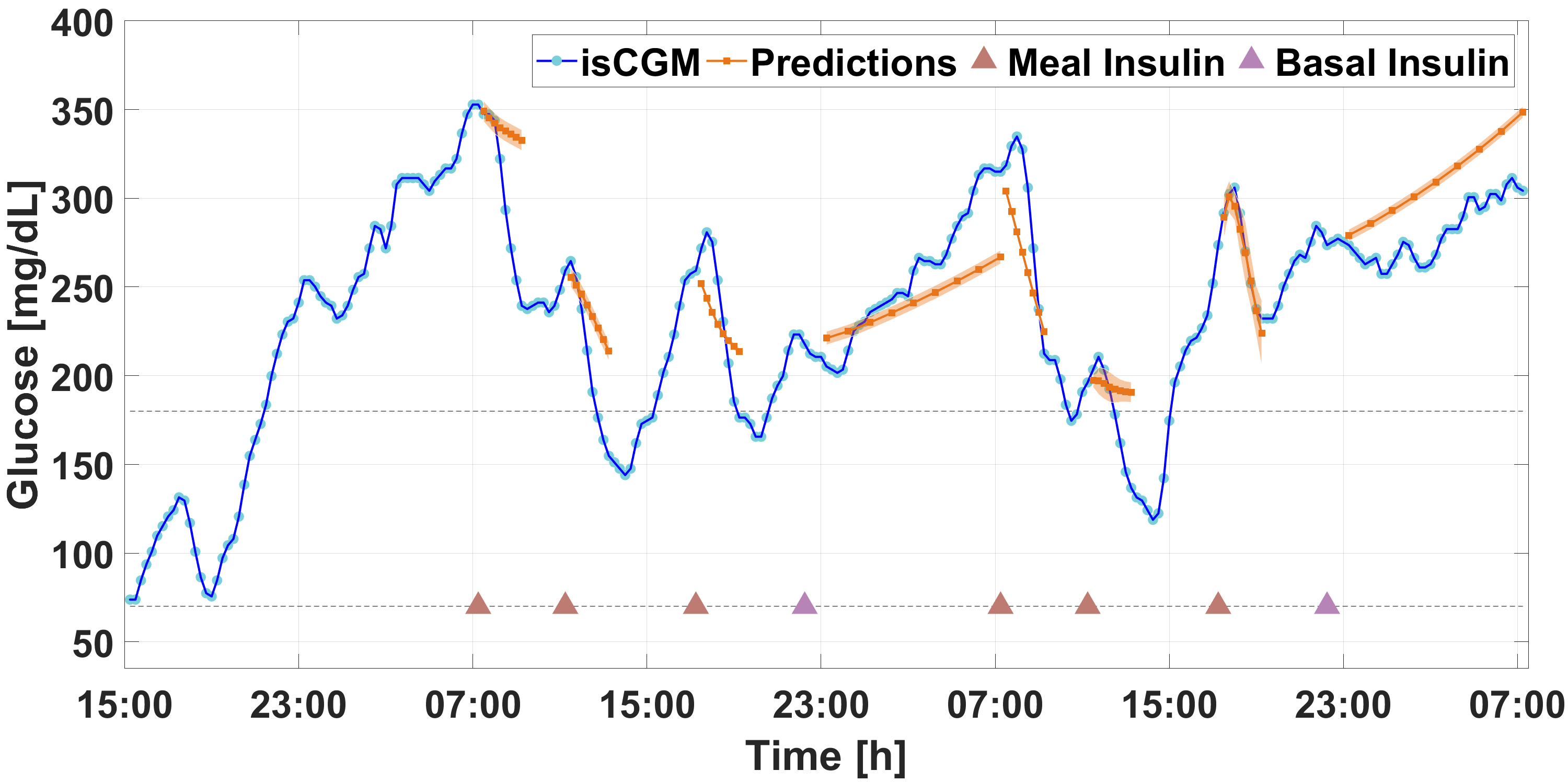}}
		\subfigure[A patient from the T2DM dataset]
		{\includegraphics[width=0.49\textwidth]{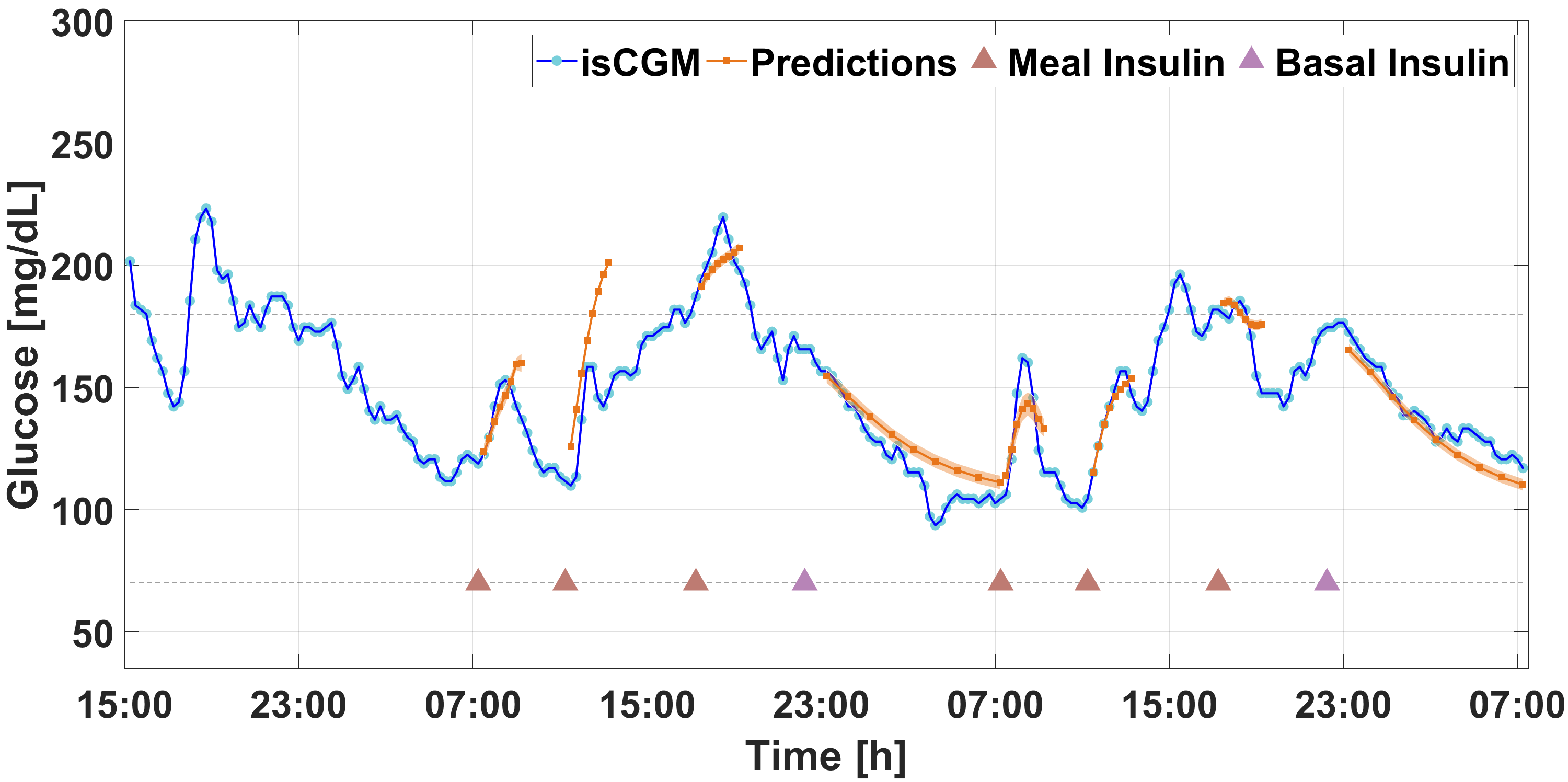}}
		\caption{Glucose predictions during the 2-day MDI decisions for the T1DM and T2DM patients.}
		\label{uob}
	\end{figure*} 
A total of 40 inpatients with T1DM with 297 treatment days and 55 inpatients with T2DM with 303 treatment days are considered in the analysis. The mean age was 49.2$\pm$19.9 years and 60.5$\pm$13.6 years for patients with T1DM and T2DM, respectively. Diabetes duration was 12.4$\pm$9.7 years and 14.0$\pm$9.1 years. Glycated hemoglobin A1c (HbA1c) was 9.3$\pm$1.9\% and 10.2$\pm$1.8\%. {\color{v1}The detailed profiles of the considered T2DM and T1DM patients can be found in Table~\ref{T5}.}

We first evaluated the prediction performance of the physiology-informed (PI) GP model during the insulin recommendations on both the T2DM and T1DM datasets. Even though there are small samples used in the training of the prediction model, the predicted glucose trajectories follow the actual glucose patterns after meals or night sleep well in both datasets (Fig.~\ref{uob}). For the population-level results of glucose prediction in the T2DM dataset, the PI-GP model achieved an average MAE of 28.4 (95\% CI: 26.8, 29.9) mg/dL vs. 30.1 (95\% CI: 28.0, 32.2) mg/dL for night glucose trajectory after basal insulin, and an average MAE of 34.2 (95\% CI: 33.0, 35.4) mg/dL vs. 41.3 (95\% CI: 39.4, 43.0) mg/dL for postprandial glucose trajectory. Similar improved results were obtained in the T1DM dataset. Besides, the online PI-GP model achieved a nearly 100\% ratio of physiologically sound prediction for both the T1DM and T2DM datasets, whereas only ratios of 67.8\% and 59.4\% were achieved by plain GP, respectively.

We further evaluated the performance of the proposed method for personalized insulin recommendations in MDI therapy. The median (min, max) run times of the method in the T1DM and T2DM datasets were 4 (0.75, 9.5) and 2 (0.25, 7.25) days, respectively, and the parallel insulin decision times were 16 (3, 38) and 8 (1, 29) times. Note that the method was first trained using the isCGM data and insulin dosage collected in the first two days of the run time, and started to provide insulin dosage recommendations in the third day. Taking the dosage decided by the physicians as references, the method achieved comparable decision performance, with low MAEs of 1.24 U (95\% CI: 1.15, 1.34) and 1.46 U (95\% CI: 1.29, 1.62) on the T1DM dataset for bolus and basal decisions, respectively (see Table~\ref{T7}). The percentages of ``clinical agreement'' (defined as dose difference $\leq$ 20\%) were 65.1\% and 83.6\% for bolus and basal decision, respectively, and the corresponding percentages of ``identical agreement'' (defined as same dosage) were 28.0\% and 18.2\%. When evaluated on the T2DM dataset, the method also exhibited a low MAE of 1.15 U (95\% CI: 1.03, 1.27) for the bolus decision, together with a ``clinical agreement'' percentage of 73.2\%, and an ``identical agreement'' percentage of 34.1\%. Similar results were observed for the basal decision (see Table~\ref{T7}). These imply that the proposed method can recommend suitable insulin dosages in the clinical scenarios based on the framework of prediction-based optimization.

\subsection{Comparison of insulin decision}
\begin{figure*}[t]
	\centering
	\centering
	\subfigure[T1DM example 1]
	{\includegraphics[width=0.32\textwidth]{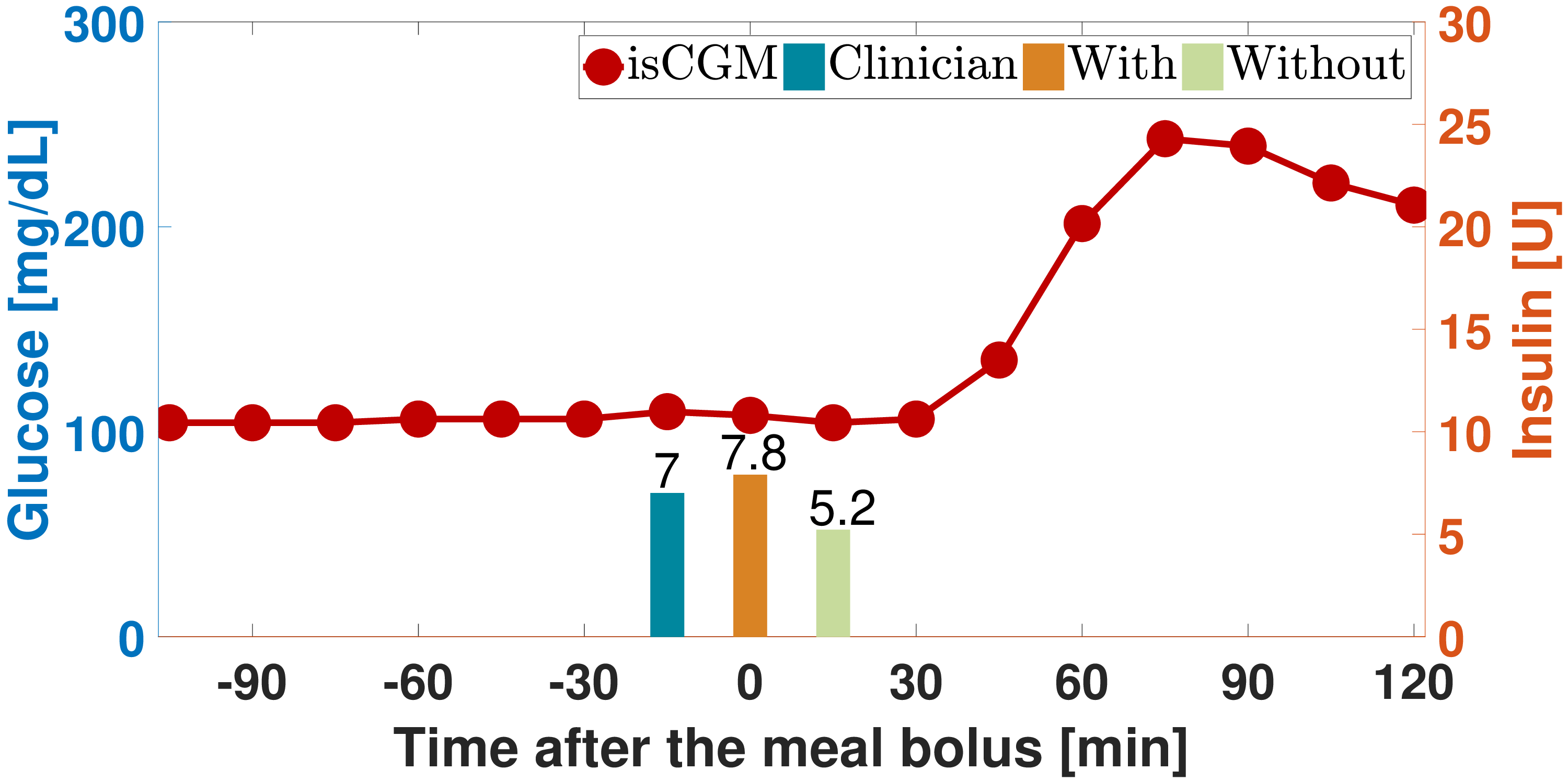}\label{e1}}
	\subfigure[T1DM example 2]
	{\includegraphics[width=0.32\textwidth]{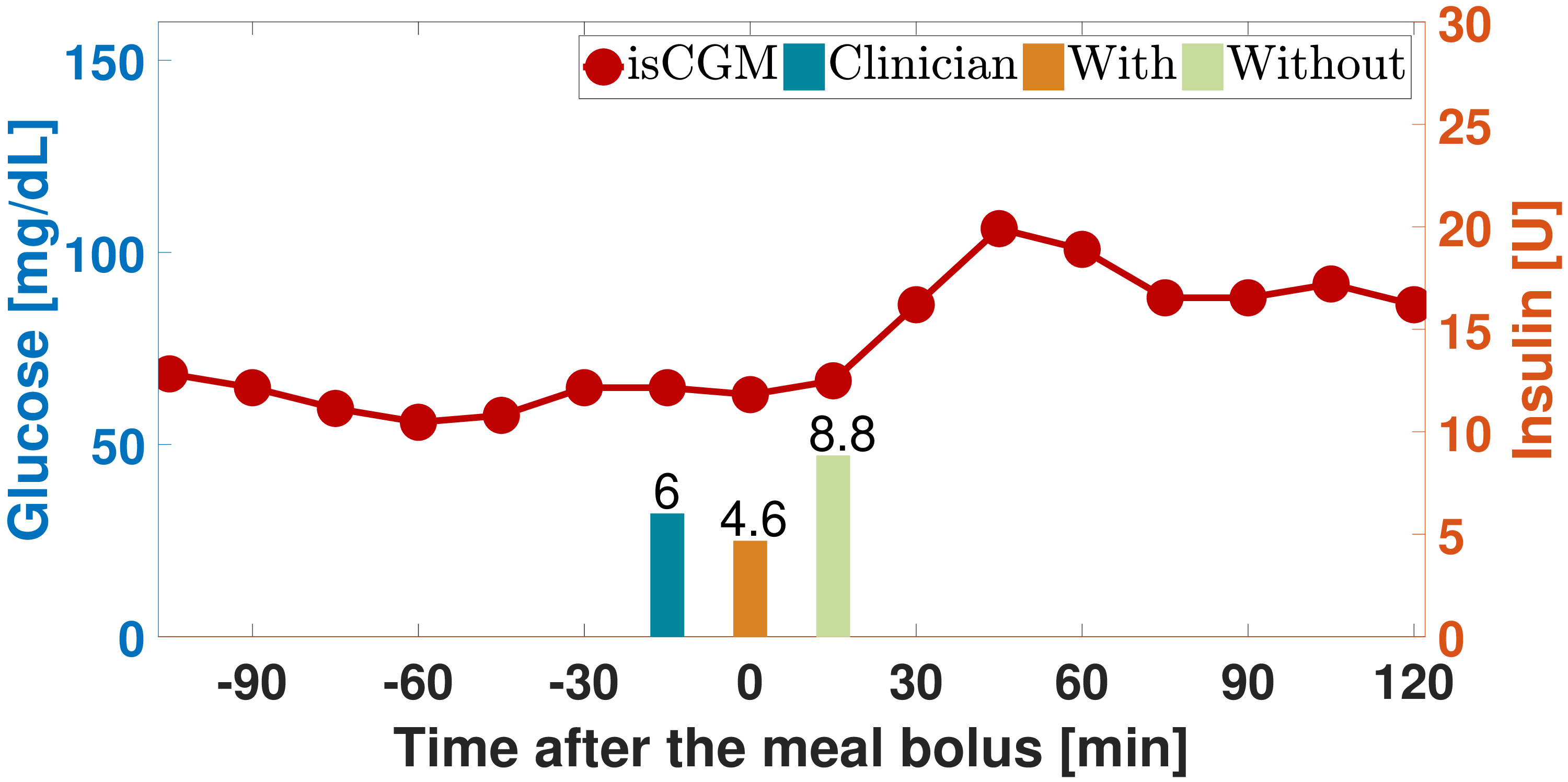}}\label{e2}
	\subfigure[T1DM example 3]
	{\includegraphics[width=0.32\textwidth]{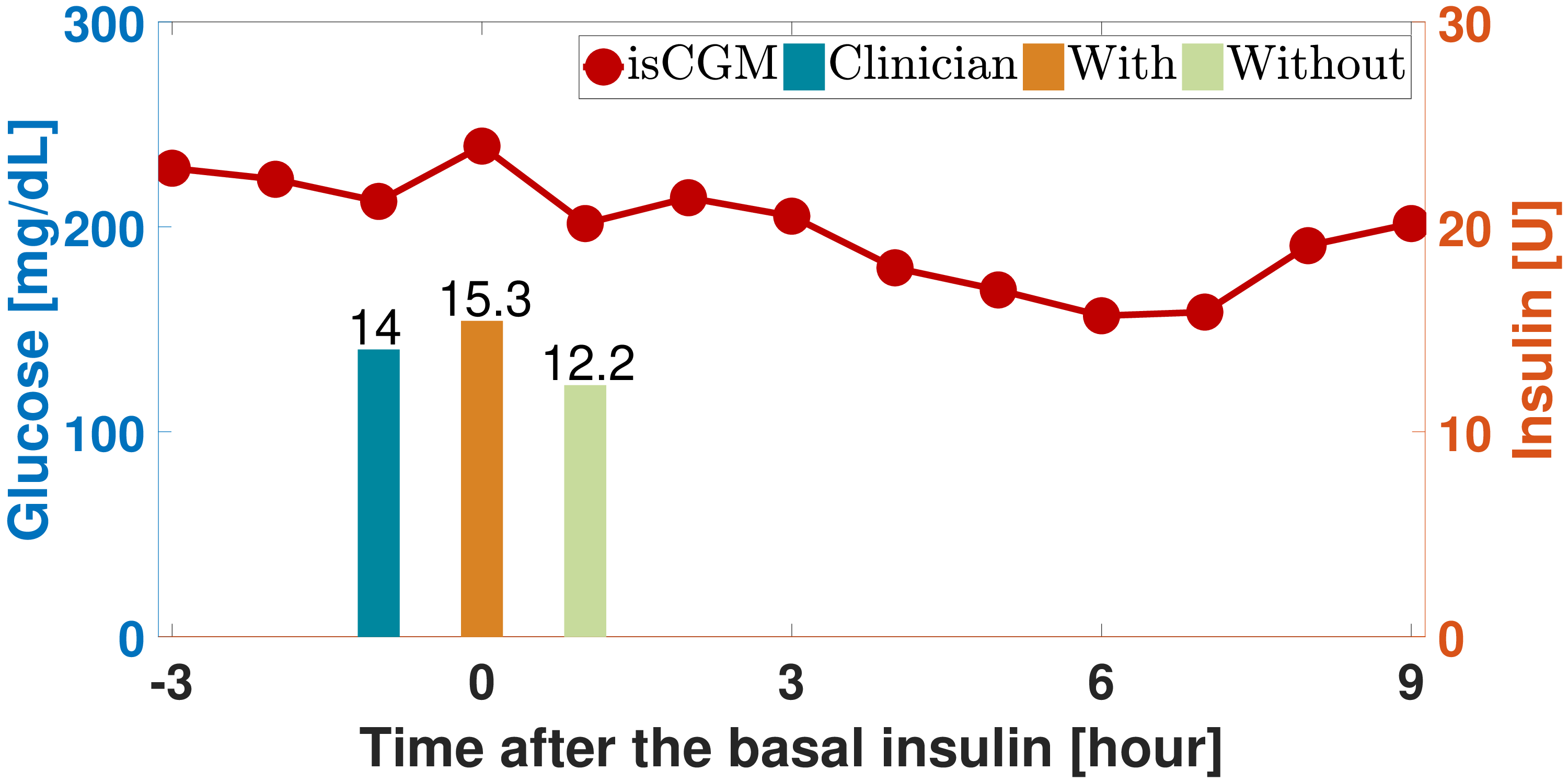}}\label{e3}
	\subfigure[T2DM example 1]
	{\includegraphics[width=0.32\textwidth]{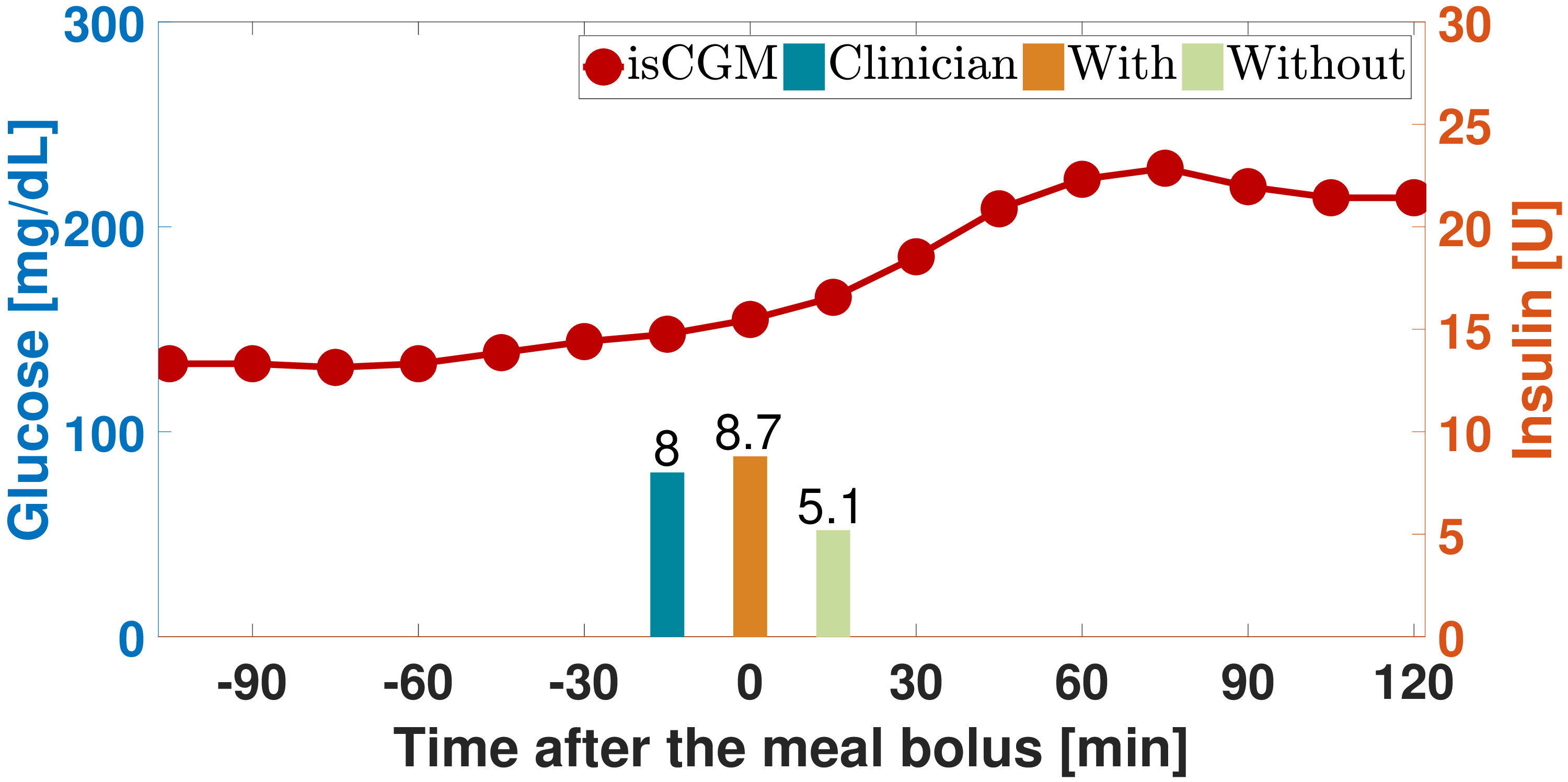}\label{e4}}
	\subfigure[T2DM example 2]
	{\includegraphics[width=0.32\textwidth]{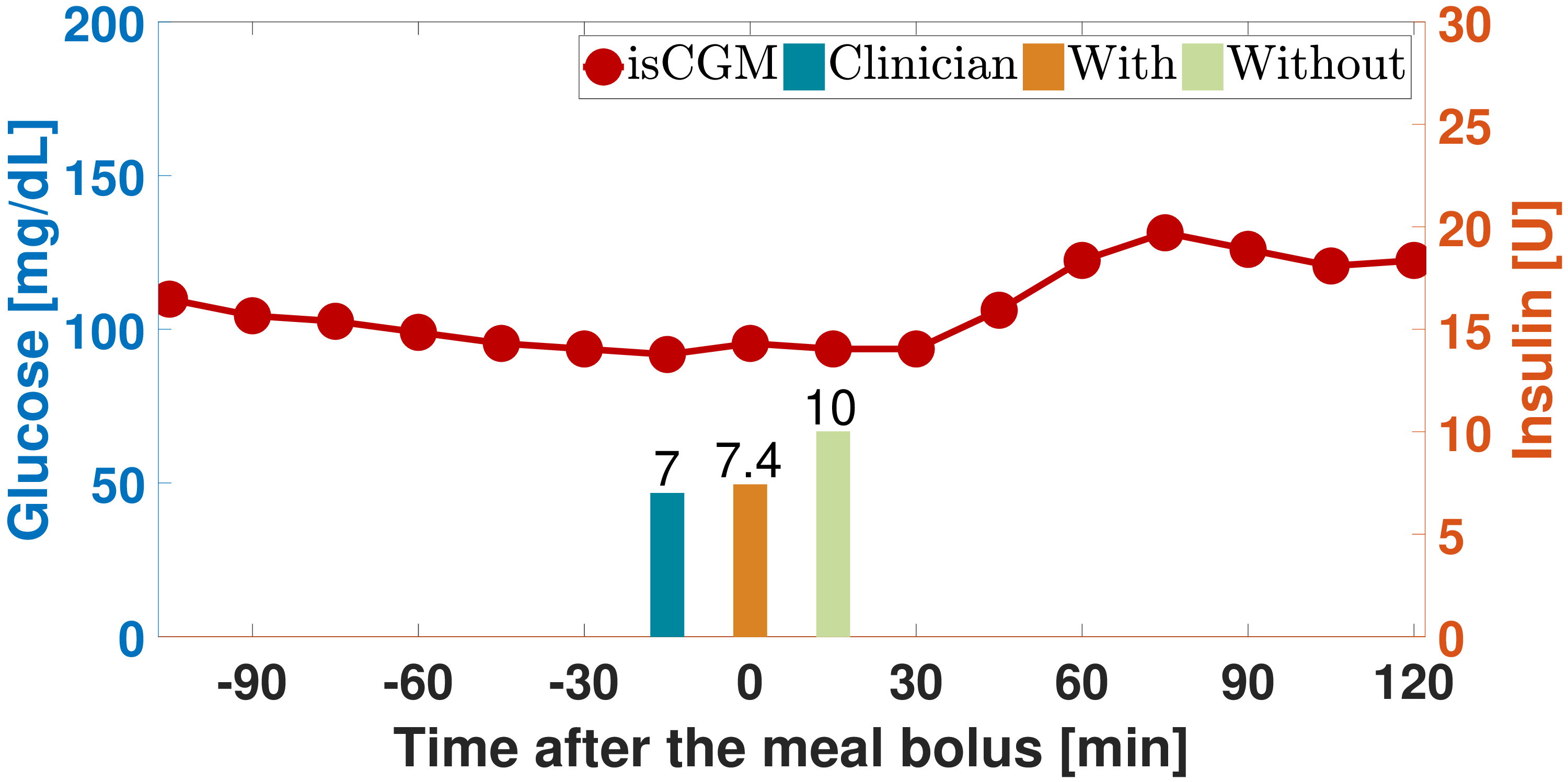}}\label{e5}
	\subfigure[T2DM example 3]
	{\includegraphics[width=0.32\textwidth]{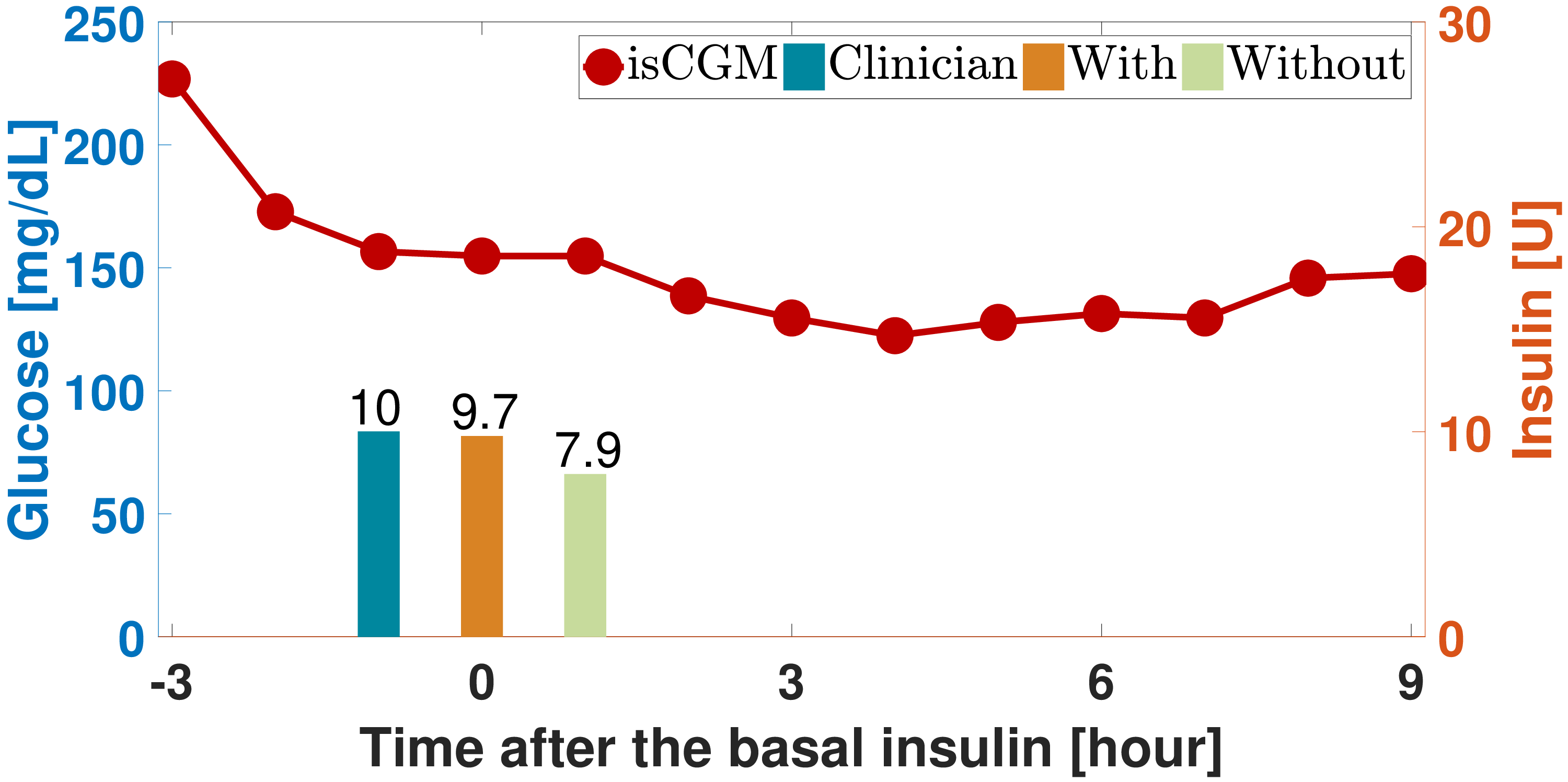}}\label{e6}
	\caption{Comparison of insulin decision for the proposed method with/without utilizing glucose metabolism priors in modeling.}
	\label{examples}
\end{figure*}

{\color{v1}In this subsection, several examples for meal bolus and basal insulin decisions from both the T1DM and T2DM dataset are provided to show the effectiveness of the proposed method equipped with physiology-informed prediction model (see Section E for a detailed introduction to the dataset). As shown in Fig.~\ref{examples}, compared with the insulin doses (denoted as blue bars) provided by clinicians, the proposed method with physiology-informed prediction model can determine appropriate dosages (denoted as orange bars) for the MDI therapy. In contrast, the method without utilizing glucose metabolism priors in modeling adopts the unacceptable decision actions (denoted as green bars). Specifically, we observe that the method with physiology-informed prediction model would recommend additional or reduced 1-2 units of the insulin dosage corresponding to the elevated and lower glucose levels (see Figs.~\ref{e1}-\ref{e4}). This reasonable increase or decrease dosage would help reduce the later happened hyperglycemia or the risk of hypoglycemia. On the contrary, the method without utilizing glucose metabolism priors in modeling tends to adjust insulin dosage in a wrong direction, which may put the patient in dangerous situations. }

\section{Conclusion}
In this work, a physiology-informed GP learning-based asymmetric risk-sensitive control method is developed for personalized MDI decisions. To provide credible glucose predictions for the decision, a physiology-constrained Gaussian process learning method is developed for glycemic dynamics modeling by encoding the physiological priors as constraints in the model training. Besides, through incorporating a penalty term associated with the previous dosage and enforcing asymmetric penalties on hyper- and hypoglycemia in the objective function, the proposed method can quickly determine suitable insulin dosage based on the feedback of the previous decision. The effectiveness and robustness of the method are verified utilizing the 10-adult cohort of the UVA/Padova simulator. Also, an advisory-mode analysis to evaluate the potential practical application in real clinical settings is performed based on the clinical data from both the T1DM and T2DM subjects. 


\section*{Acknowledgment}
Access to the distributed version of the University of Virginia (UVA)/Padova metabolic simulator for research purposes is acknowledged.

\appendices 
\section{Physiology-informed constraints for hyperparamter optimization}
In this appendix, the specific expressions of the constraints in \eqref{constrains} for the case of the mean and covariance functions in \eqref{mean}-\eqref{SE} are provided. Given the training inputs $X_k$ and outputs $Y_k$, the predictive mean $y_*$ has the following form:
\begin{align}
y_*=m(x_*)+K_*^\top(K+\sigma^2_\omega I)^{-1}(Y_k-m(X_k)). 
\end{align}
Since the GP regression is a linear smoother, the predictive mean can be re-written as \cite{Rasmussen2006}:
\begin{align}
y_*=a^\top x_*+b+ \sum_{i=1}^N \alpha_i k(x_*, x_i), 
\end{align}
where $m(x_*) = a^\top x_*+b$ is used; $N$ is the sample number and $\alpha_i$ is the $i$th element of $\alpha := (Y_k-m(X_k))^{\top}(K+\sigma^2_\omega I)^{-1}$. We can then calculate $\partial y_*/\partial u_*$ as 
\begin{align}
\frac{\partial y_*}{\partial u_*} &=a_{u} + \sum_{i=1}^N \alpha_i \frac{\partial k(x_*, x_i)}{\partial u_*}, \\
&=a_{u} - \sum_{i=1}^N \alpha_i k(x_*, x_i) \frac{u_*-u_i}{l_{m+1}^2},
\end{align}
where $k(x_*, x_i) = \sigma_f^2\exp\left(-\frac{1}{2}(x_*-x_i)^\top\Omega^{-1}(x_*-x_i)\right)$ is used and $a_u$ is the element of $a$ corresponding to $u$. Defining $S(x_*, x_i) := k(x_*, x_i) (u_*-u_i)/l_{m+1}^2$, and according to $\partial y_*/\partial u_* < 0$, we have
\begin{align}
a_{u} &< \sum_{i=1}^N \alpha_i S(x_*, x_i), ~\forall u_* \in [0, u_{\max}],\\
&<-\sqrt{\sum_{i=1}^N \alpha_i^2 \sum_{i=1}^N S(x_*, x_i)^2},  ~\forall u_* \in [0, u_{\max}], \label{Cauchy}
\end{align}
where the Cauchy–Schwarz inequality is applied to obtain the inequality \eqref{Cauchy}. We can then easily compute the maximum value of $S(x_*, x_i)$ for $u_*$ using its partial derivative to $u_*$, and a conservative result for \eqref{Cauchy} is further obtained based on the maximum value:
\begin{align}
a_u &<-\sqrt{\sum_{i=1}^N \alpha_i^2\sum_{i=1}^N (\sigma_f^2\frac{\exp(-\frac{1}{2}+c_i)}{|l_{m+1}|})^2}, \\
&<-\sqrt{\sum_{i=1}^N \alpha_i^2\sum_{i=1}^N (\sigma_f^2\frac{\exp(-\frac{1}{2})}{|l_{m+1}|})^2},   
\end{align}
where $c_i := -\frac{1}{2}(x_{*/u}-x_{i/u})^\top\Omega_{/u}^{-1}(x_{*/u}-x_{i/u})$ excluding the element $u$. Following the similar arguments, we can also obtain the constraint for $\partial y_*/\partial d_* > 0$ as
\begin{align}
a_d > \sqrt{\sum_{i=1}^N \alpha_i^2\sum_{i=1}^N (\sigma_f^2\frac{\exp(-\frac{1}{2})}{|l_{m+2}|})^2}. 
\end{align}

\bibliographystyle{IEEEtran}
\bibliography{Caidh}

\end{document}